# Sincronização de Disparos em Redes Neuronais com Plasticidade Sináptica

# (Spikes Synchronization in Neural Networks with Synaptic Plasticity)

Rafael R. Borges[1]; Kelly C. Iarosz[2*]; Antonio M. Batista[3]; Iberê L. Caldas[4]; Fernando S. Borges[1]; Ewandson L. Lameu[1]

1 Pós-Graduação em Ciências/Física, Universidade Estadual de Ponta Grossa, Paraná, Brasil.
2 Instituto de Física (Pós-Doutorado CNPq), Universidade de São Paulo, São Paulo, Brasil.
3 Departamento de Matemática e Estatística, Universidade Estadual de Ponta Grossa, Paraná, Brasil.
4 Departamento de Física Aplicada, Instituto de Física, Universidade de São Paulo, São Paulo, Brasil.
*e-mail: kiarosz@gmail.com

**Resumo**

Neste artigo, investigamos a sincronização de disparos neuronais em uma rede neuronal com plasticidade sináptica e perturbação externa. Nas simulações, a dinâmica neuronal é descrita pelo modelo de Hodgkin-Huxley, considerando sinapses químicas (excitatórias) entre neurônios. De acordo com a sincronização dos disparos é esperado que uma perturbação cause regimes não sincronizados. No entanto, na literatura existem trabalhos que mostram que a combinação de plasticidade sináptica e perturbação externa podem gerar regime sincronizado. Este artigo descreve o efeito da plasticidade sináptica na sincronização, onde consideramos a perturbação descrita por uma distribuição uniforme. Este estudo é relevante para pesquisas de controle de distúrbios neuronais.

**Palavras-chave:** Neurônios, plasticidade, sinapses químicas.

**Abstract**

In this paper, we investigated the neural spikes synchronisation in a neural network with synaptic plasticity and external perturbation. In the simulations the neural dynamics is described by the Hodgkin Huxley model considering chemical synapses (excitatory) among neurons. According to neural spikes synchronisation is expected that a perturbation produce non synchronised regimes. However, in the literature there are works showing that the combination of synaptic plasticity and external perturbation may generate synchronised regime. This article describes the effect of the synaptic plasticity on the synchronisation, where we consider a perturbation with a uniform distribution. This study is relevant to researches of neural disorders control.

## 1 Introdução

O crescente número de diagnósticos de perturbações neurológicas é uma preocupação de nível mundial, tanto em termos do bem estar humano como no impacto na economia. Tais perturbações incluem além de traumatismos cerebrais, infecções neurológicas, esclerose múltipla, a doença de Parkinson[1], epilepsias e Mal de Alzheimer. Além de afetarem pessoas de todos os países, independentemente da idade, sexo, nível de educação ou rendimento, tais doenças também afetam o Produto Interno Bruto (PIB) dos países, através da falta de produtividade que decorre do desenvolvimento de uma doença mental, estima-se que essa quebra no PIB seja de 4% ao ano [1].

Segundo relatório publicado pela Organização Mundial de Saúde [1], estima-se que, todos os anos, 6,9 milhões de pessoas morrem devido a perturbações neurológicas. Cerca de 1% da população com mais de 65 anos apresenta a doença de Parkinson, isso corresponde a mais de quatro milhões de pessoas afetadas. E as previsões até 2020 são assustadoras, mais de 85 milhões de pessoas sofrerão com doenças neurológicas, sendo que destas, 50 milhões terão epilepsia[2] e mais de 35 milhões terão Alzheimer[3] ou outras demências [1].

Uma das tentativas para solucionar ou pelo menos amenizar essa situação alarmante é o investimento em pesquisas científicas envolvendo diversas áreas e formas de atacar os problemas identificados. Estes que envolvem neurociências funcionam como quebra-cabeças, ou seja, são separados em muitas peças e cada uma é tão necessária quanto a outra, e a física tem colaborado muito para o avanço de modelos que podem auxiliar em tratamentos e desenvolvimento de mecanismos para uma melhor qualidade de vida. Conceitos envolvendo oscilações, circuitos e equações diferenciais permitem a interpretação acurada de fenômenos biológicos, dando cunho para desenvolvimento de pesquisas importantes.

Neste trabalho abordamos a sincronização de neurônios utilizando o modelo de Hodgkin-Huxley (HH) [2], com medidas realizadas a partir dos picos de disparos (do inglês *spikes*) [3, 4] quando submetidos a perturbações externas e plasticidade sináptica. Este estudo é uma aplicação dos conhecimentos em física e tem efeito

---

[1] Doença progressiva do movimento devido à disfunção dos neurônios que controlam e ajustam a transmissão dos comandos conscientes vindos do córtex cerebral para os músculos do corpo humano.

[2] Grupo de transtornos neurológicos de longa duração caracterizados por atividade excessiva e anormal das células nervosas do córtex cerebral.

[3] Doença neurodegenerativa que provoca declínio de funções intelectuais, reduz a capacidade de trabalho, relação social e interfere no comportamento e na personalidade.

relevante na descoberta de novas formas de controle de patologias.

A sincronização[4] está diretamente ligada ao ajuste dos ritmos de osciladores[5] devido à sua interação [5]. Pode ocorrer com qualquer tipo de oscilador, seja ele físico, químico e biológico [3]. O fenômeno de sincronização é foco de estudo em muitas áreas, pois é frequentemente observado na natureza em sistemas acoplados com comportamento periódico ou caótico [5].

Conhecendo os princípios que resultaram na sincronia ou na assincronia de um sistema, é possível responder questões referentes à engenharia, matemática, física e neurociência entre outras. A sincronização de oscilações neuronais, por exemplo, pode ser observada desde os primeiros estágios de muitos sistemas sensoriais de animais, como insetos, sapos e primatas. Alguns estudos realizados com mamíferos, demonstraram que as sinapses[6] entre os neurônios são estabelecidas nos primeiros estágios do desenvolvimento e a sincronização das oscilações neuronais podem ocorrer durante o desenvolvimento e aprendizagem [6].

Como o fenômeno de sincronização mostra a correlação da dinâmica de diferentes sistemas quando estes apresentam alguma forma de interação, seu papel na transmissão de informação e estudo de enfermidades é cada vez mais aplicado [7, 8]. Em pesquisas recentes sobre doença de Alzheimer, análises baseadas na sincronização e dessincronização neuronal são consideradas e auxiliam na conclusão de que a dinâmica de oscilações do tálamo e do córtex são significativamente influenciadas pela perda sináptica cortico-cortical [9].

Em patologias como epilepsia e doença de Parkinson [10], fisiologias do funcionamento motor e processamento de informações sensoriais, estudos mostram que é benéfico conhecer a sincronia das oscilações neuronais e como suprimi-las afim de minimizar as crises [11, 12].

Uma rede pode sofrer perturbações, ou seja, alteração nas condições normais em que se encontra. No caso da rede neuronal, essa perturbação externa pode ser causado como algum dos cinco sentidos (audição, olfato, paladar, tato e visão). Tais perturbações são causadas pela inserção de uma perturbação, que pode variar em termos de intensidade.

A plasticidade sináptica é um fenômeno neuronal, postulado no final da década de 1940 pelo psicólogo Canadense Donald Hebb [13]. Ele acreditava que se dois neurônios apresentassem uma atividade conjunta, a intensidade das sinapses entre eles deveria ser reforçada. Essa proposta de Hebb foi comprovada experimentalmente por Bliss e colaboradores, realizando medidas com neurônios do hipocampo do cérebro de coelhos [14, 15].

Entre os diferentes tipos de processos sinápticos descritos na literatura, abordaremos neste artigo a plasticidade sináptica dependente do tempo de disparos de entrada e saída dos neurônios (do inglês spike timing dependent-plasticity: STDP). A inserção deste tipo de regra de plasticidade, nos permitirá ajustar a intensidade das conexões, propiciando um regime de disparos neuronais em tempos próximos, o que contribui para a sincronização dos neurônios [16].

A seguir apresentaremos a metodologia aplicada, o desenvolvimento, os resultados e conclusões referentes à investigação da sincronização dos disparos em uma rede de neurônios com sinapses modificáveis via STDP com e sem a presença de perturbação.

## 2 Metodologia

Desenvolvemos o presente trabalho utilizando ferramentas computacionais e dados de trabalhos experimentais obtidos na literatura. Para a elaboração de rotinas computacionais, utilizamos a Linguagem C [17].

Na fase de elaboração de rotina, a escolha do modelo neuronal adequado foi um passo importante, que depende do objetivo da análise a ser efetuada. Os principais modelos que encontramos na literatura, foram os do tipo integra-dispara [18-20], Izhikevich [21, 22], FitzHugh-Nagumo [23], Hindmarsh-Rose [24], Morris-Lecar [25], Wilson [26] e Hodgkin-Huxley [2]. Comparamos os modelos, e chegamos a conclusão de que o modelo dos pesquisadores Hodgkin e Huxley nos ofereceria as propriedades de disparos, rajadas e uma interpretação biológica significativa.

Após a simulação do primeiro neurônio, e com este respondendo corretamente, passamos a montar a rede com os demais neurônios. Conectamos os neurônios de forma global, pois com esta forma podemos simular sistemas que possuem ligações de longo alcance que ocorrem em física e biologia [27]. No acoplamento global, também conhecido como acoplamento de campo médio [28], um neurônio pode interagir com todos os outros neurônios da rede.

Após o acoplamento ser inserido e testado na rotina, passamos a inserção de uma perturbação externa e da STDP. A perturbação externa ($I_{ext}$) é uma das ferramentas que pode ser utilizada para a supressão da sincronização neuronal [29]. Nesse trabalho, a perturbação externa foi inserida de maneira aleatória e com uma intensidade suficiente para gerar disparos neuronais. Isto causa uma alteração no tempo de disparos dos neurônios e dependendo da intensidade, propicia um regime não sincronizado.

A STDP é inserida logo em sequência. Entra no modelo modificando o termo no acoplamento e, desta forma, determina a evolução dos pesos sinápticos em função da atividade dos neurônios pré e pós-sináptico, ocasionando uma potenciação ou uma depressão destes [30].

## 3 O neurônio e seu comportamento

Um modelo neuronal procura simular o mais realisticamente possível, o que ocorre em um neurônio biológico. Quanto a um neurônio biológico, nos referimos a uma célula do sistema nervoso responsável pela condução de impulsos nervosos. Atualmente, acredita-se que no cérebro existam cerca de 86 bilhões de neurônios[7] [31, 32], cada um com suas peculiaridades. Na Figura 1, observamos

---

[4] Derivação da palavra síncrono e significa "ocorrência ao mesmo tempo".
[5] Corpo que possui uma oscilação. Oscilação é o movimento de um corpo que passa e torna a passar alternativamente pelas mesmas posições.
[6] No século 19, Santiago Ramón y Cajal deixou relatado em seus escritos e figuras, a existência de pequenos espaços entre células nervosas. Mais tarde, Charles Sherrington atribuí nome aos espaços que estavam ligados a passagem de informação: sinapses.

[7] Nas referências [31-33] é possível conhecer mais a fundo sobre as células neuronais em geral e suas especificidades.

uma representação esquemática de um neurônio biológico com seus componentes básicos.

O soma é a parte central do neurônio, nele estão presentes as organelas e o núcleo. Os dendritos são numerosos prolongamentos que recebem estímulos nervosos e os transmite para o corpo da célula e o axônio é responsável pela condução dos impulsos elétricos que partem do corpo celular, até outro local mais distante. Os neurônios variam em forma, tamanho e conexões [32].

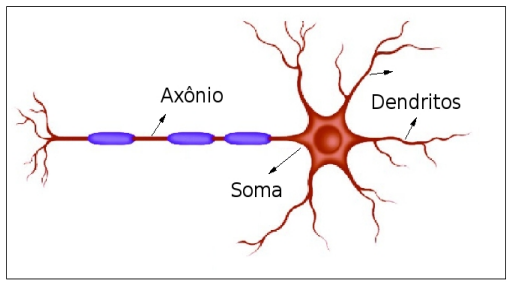

**Figura 1:** Componentes básicos de um neurônio.

Alguns neurônios podem responder a estímulos disparando potenciais de ação a uma frequência constante, outros podem não apresentar uma taxa de disparos constante, então disparam no início do estímulo e, logo na sequência diminuem a taxa de disparos mesmo que o estímulo ainda permaneça.

Na Figura 2, observamos a atividade neuronal caracterizada por um disparo neuronal com amplitude máxima para uma corrente inicial de 20 mV [33]. Este tipo de comportamento é característico da evolução temporal do potencial de ação da membrana de algumas células neuronais [31]. Na ausência de uma corrente (uma perturbação ou estímulo externo), o potencial permanece constante. Entretanto, dependendo do estímulo fornecido à célula, o potencial da membrana começa variar, até que atinge um limiar. A partir desse limiar a subida é rápida até atingir a amplitude máxima[8]. O retorno ao potencial de repouso[9] é mais lento que a ascensão ao pico e, durante o intervalo de quando o potencial retorna ao valor do limiar até alcançar novamente o potencial de repouso[10], temos um período refratário, onde o neurônio fica quiescente até poder disparar novamente [3].

Os valores de potencial sofrem modificações devido às junções (fendas) conhecidas como sinapses. Nas sinapses, um neurônio influencia diretamente outro(s) neurônio(s), através da transmissão de sinais elétricos ou químicos. O neurônio que ocasiona a alteração é chamado pré-sináptico, enquanto o neurônio que sofreu a ação é chamado pós-sináptico, conforme indicado na Figura 3.

### 3.1 Sinapses

As sinapses foram observadas na década de 50, mas precisamente em 1959, quando Edward Gray publicou fotomicrografias da fenda sináptica [34]. Tal comprovação mostra que as sinapses são estruturas microscópicas de contato entre neurônios e/ou outras células, ou seja, são junções especializadas [31, 33]. Podem ser elétricas ou químicas. As sinapses elétricas apresentam transmissão rápida de impulsos elétricos, porém, não processam informações. São sincronizadoras de atividade neuronal. Já as sinapses químicas, são moduláveis, podem transmitir e modificar informações e manifestam-se na forma excitatória, resultando em um novo impulso nervoso (como a utilizada neste trabalho) ou na forma inibitória, impedindo a passagem de impulsos [33, 35].

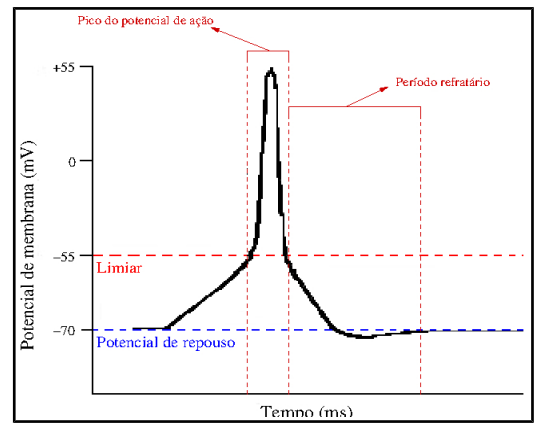

**Figura 2:** Representação da atividade de disparos neuronais, conforme o potencial de membrana sofre alterações, observamos o processo de construção de um pico de potencial de ação neuronal.

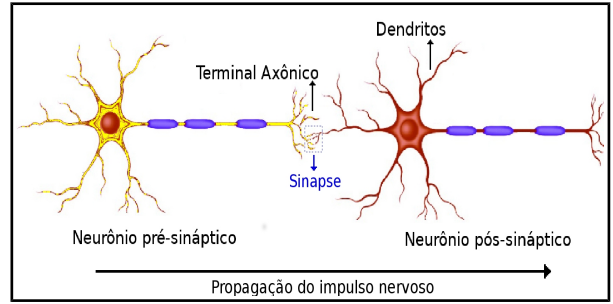

**Figura 3:** Neurônios pré e pós-sinápticos, demonstrando a região de acoplamento onde ocorre a sinapse. É possível verificar o sentido de propagação do sinal elétrico entre os neurônios pré e pós-sinápticos.

### 3.2 Plasticidade sináptica dependente do tempo de disparos (do inglês: *Spike timing dependent plasticity-STDP*).

Donald Hebb formulou uma hipótese em 1949 sobre conexões neuronais sendo reforçadas quando os neurônios apresentam uma atividade conjunta. Baseado nessa hipótese e nas descobertas experimentais posteriores, Wulfram Gerstner[11] mostrou que a comunicação entre os neurônios depende da relação entre os instantes dos disparos dos potenciais de ação [36-38]. A proposta de Gerstner foi comprovada experimentalmente em 1997 por Henry Markram [36] e em 1998 por Guo-qiang Bi e Mu-ming Poo [16].

---

[8] Esse processo é chamada despolarização da membrana.
[9] A queda do pico até o limiar é conhecida como repolarização.
[10] Período em que o potencial permanece abaixo do potencial de repouso é chamado hiperpolarização.

[11] Professor Wulfram Gerstner é diretor do Laboratótio de Neurociência Computacional (LCN) na Escola Politécnica Federal de Lausana na Suíça.

Na figura 4 é apresentada uma ilustração do resultado obtido experimentalmente por Bi e Poo [16]. Ocorre fortalecimento (potenciação) na intensidade das conexões entre neurônios se o neurônio pré-sináptico dispara antes do pós-sináptico (Δt>0). Caso contrário, (Δt<0), ocorre o enfraquecimento (depressão) nas conexões. Este processo corrobora o Postulado de Hebb, pois a atividade conjunta dos neurônios leva a modificações nas intensidades das sinapses.

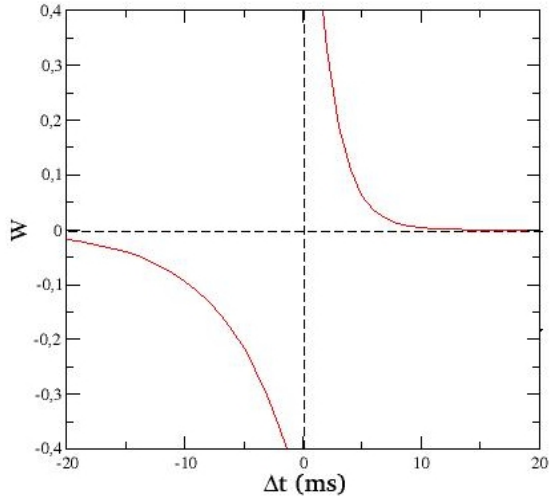

**Figura 4:** A representação apresentada, mostra a plasticidade dependente do tempo dos disparos neuronais (STDP). Podemos observar a variação da intensidade de conexão entre os neurônios da rede (W) e sua dependência com os intervalos de disparos dos neurônios pré e pós-sinápticos.

## 4 O modelo

Por meio de modelos é possível realizar simulações e estudos comportamentais de neurônios. Os modelos capacitivos representam de forma satisfatória algumas das variáveis de um neurônio. Logicamente, não é um modelo perfeito, porque ainda não se conhece um modelo que seja capaz de simular totalmente as funções de células tão complexas.

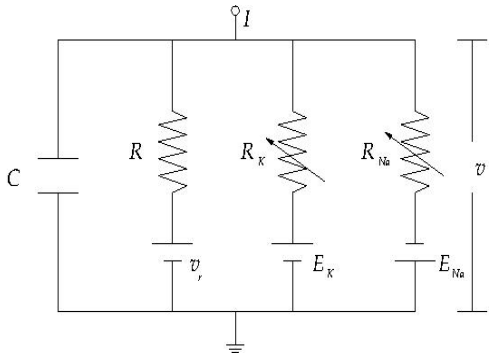

**Figura 5:** Circuito elétrico capacitivo, onde a membrana neuronal é representada por um capacitor de placas paralelas, e alguns possíveis canais iônicos são apresentados como ramos do circuito.

O modelo proposto nas simulações é um circuito capacitivo, representando uma membrana neuronal e os canais iônicos que por ela passam. A Figura 5 é uma representação esquemática do circuito considerado neste artigo, com um capacitor de placas paralelas, com capacitância $\mathcal{C}$. Em paralelo ao capacitor, encontram-se a associação de três ramos, cada um equivalente a um canal iônico. Os canais de Sódio (Na) e de Potássio (K) são representados por potenciômetros que variam de acordo com o potencial $\mathcal{V}$ e fontes elétricas ($\mathcal{E}_{Na}$, $\mathcal{E}_K$) representando o potencial reverso (dado pela equação de Nernst [32, 33]), enquanto o canal que representa os demais íons é representado por um resistor $\mathcal{R}$ conectado em série com uma fonte $\mathcal{V}_r$. O circuito tem uma corrente de entrada $I$ que em um neurônio seriam sinapses, ou outro sinal externo. Este circuito foi proposto por volta de 1952 por Alan Lloyd Hodgkin e Andrew Fielding Huxley, os experimentos eram voltados ao estudo de potenciais de ação, e eram realizados em um axônio de grosso calibre de um molusco (*Loligo Pealei*) [2]. E pode ser resolvido utilizando as Leis de Kirchhoff.

O estudo dos pesquisadores Hodgkin e Huxley foi possível graças ao avanço dos mecanismos instrumentais da década de 50. Por meio do conjunto de equações diferenciais não-lineares de 1ª ordem, e parâmetros coletados experimentalmente, o modelo descreveu e ainda descreve o comportamento de células neuronais. Na tabela 1, apresentamos um apanhado geral das medidas experimentais utilizadas para simulações envolvendo o modelo HH [2].

**Tabela 1:** Parâmetros para o modelo de Hodgkin-Huxley considerando o potencial de repouso igual a −65 mV.

| Sigla | Significados | Valores |
|---|---|---|
| $C_M$ | Capacitância da membrana | 1 μF/cm² |
| $V_{Na}$ | Potencial reverso (Sódio) | 50 mV |
| $V_K$ | Potencial reverso (Potássio) | -77 mV |
| $V_l$ | Potencial reverso (demais íons) | -54,4 mV |
| $g_{Na}$ | Condutância (íons sódicos) | 120 mS/cm² |
| $g_K$ | Condutância (íons potássicos) | 36 mS/cm² |
| $g_l$ | Condutância (demais íons) | 0,3 mS/cm² |
| $I$ | Corrente inicial | 9,0 – 10,0 μA/cm² |

Fonte: [39].

O circuito da Figura 5, ajustado aos parâmetros apresentados na Tabela 1 é descrito pelo sistema de equações diferenciais de 1ª ordem:

$$C_M \dot{V} = I - g_K n^4 (V - V_K) - g_{Na} m^3 h (V - V_{Na}) - g_l (V - V_l),$$
$$\dot{m} = \alpha_m (1 - m) - \beta_m m,$$
$$\dot{h} = \alpha_h (1 - h) - \beta_h h,  \quad (1)$$
$$\dot{n} = \alpha_n (1 - n) - \beta_n n,$$

onde $C_M$ é a capacitância da membrana neuronal, $V$ é o potencial de membrana do neurônio, $V_K$, $V_{Na}$ e $V_l$, são os potenciais devido aos íons de Potássio, Sódio e demais íons presentes no fluído celular[12], respectivamente. As condutâncias específicas máximas para cada um dos canais iônicos são $g_K$, $g_{Na}$ e $g_l$. A corrente constante que desencadeia uma sequência periódica de potenciais de ação[13] neuronais é representada por $I$. As variáveis $m$ e $h$ controlam os estados onde a membrana conduz ou não a corrente, referente aos íons de Sódio. Enquanto *n* comanda a fração de canais de Potássio. As funções *α* e *β* foram determinadas experimentalmente [39] e são dadas por:

$$\alpha_n(V) = 0,01 \frac{10-V}{\exp\left(\frac{10-V}{10}\right)-1},$$

$$\alpha_m(V) = 0,1 \frac{25-V}{\exp\left(\frac{25-V}{10}\right)-1},$$

$$\alpha_h(V) = 0,07 \exp\left(\frac{-V}{20}\right),$$

$$\beta_n(V) = 0,125 \exp\left(\frac{-V}{80}\right), \quad (2)$$

$$\beta_m(V) = 4\exp\left(\frac{-V}{18}\right),$$

$$\beta_h(V) = \frac{1}{\exp\left(\frac{30-V}{10}\right)+1}.$$

---
[12] É possível verificar mais a fundo a questão dos potenciais nas bibliografias [31-33].
[13] Também chamado de impulso nervoso [32].

Podemos investigar o comportamento de uma população de neurônios interligados por sinapses químicas excitatórias. Consideramos a configuração global para a rede neuronal, onde todos os neurônios estão conectados, conforme proposto no trabalho de Popovych e colaboradores [29]. Desta forma o sistema de equações diferenciais de 1ª ordem não-linear passa a ser:

$$C_M \dot{V}_i = I - g_K n_i^4(V_i - V_K) - g_{Na} m_i^3 h_i(V_i - V_{Na}) - g_l(V_i - V_l) + \frac{V_r - V_i}{N}\sum_{j=1}^N a_{ij} s_j(t) + I_{EXT},$$

$$\dot{m}_i = \alpha_m(1-m_i) - \beta_m m_i,$$
$$\dot{h}_i = \alpha_h(1-h_i) - \beta_h h_i, \quad (3)$$
$$\dot{n}_i = \alpha_n(1-n_i) - \beta_n n_i,$$
$$\dot{s}_i = \frac{5(1-s_i)}{1+\exp(-(V_i+3)/8)} - s_i,$$

onde $V_r = 20$ mV é o potencial reverso, N é o número de neurônios, $a_{ij}$ fornece a intensidade ou intensidade do acoplamento entre o neurônio pré-sináptico *j* e o pós-sináptico *i*. O termo $I_{ext}$ é a intensidade da perturbação externa aplicada aleatoriamente nos neurônios com uma duração igual a 1 ms. Cada neurônio recebe em média uma perturbação a cada 14 ms. O acoplamento sináptico entre uma população de neurônios é representado por $s_j(t)$ que representa a interação via sinapses químicas excitatórias por meio de mecanismo de integração do potencial pós-sináptico dos neurônios (PSP) [3, 34]. No caso deste estudo, é a forma de conexão existente entre ou neurônios da rede, resultante de processos de troca iônica, chamado acoplamento excitatório [29]. As intensidades das conexões podem ser controladas por processos biológicos, um desses processos é a plasticidade sináptica dependente do tempo (STDP). A STDP é um modelo de plasticidade sináptica que leva em consideração os tempos de disparos entre os neurônios pré e pós-sinápticos.

O principal efeito do fenômeno da STDP é o acréscimo (potenciação), ou decréscimo (depressão) na intensidade das sinapses [16], isso ocorre em função da ordem temporal dos disparos, dados por:

$$W(\Delta t_{ij}) = \begin{array}{l} A_1 \exp(\Delta t_{ij}/\tau_1) \, se \, \Delta t_{ij} \geqslant 0 \\ -A_2 \exp(-\Delta t_{ij}/\tau_2) \, se \, \Delta t_{ij} < 0 \end{array}, \quad (4)$$

onde $\Delta t_{ij} = t_i - t_j$, com *i* representando o neurônio pós-sináptico e *j* o neurônio pré-sináptico. Se o neurônio *j* dispara alguns instantes antes que o neurônio *i*, $\Delta t_{ij} > 0$, e assim, ocorre a potenciação neste sentido, pois este colaborou para o disparo de *i*. Caso contrário, ocorre a depressão na sinapse entre os neurônios *i* e *j*. Os parâmetros do modelo são $A_1 = 1$; $A_2 = 0,5$; $\tau_1 = 1,8$ ms e $\tau_2 = 6,0$ ms [29]. A atualização dos pesos sinápticos é dada por

$$a_{ij} \Rightarrow a_{ij} + \delta W(\Delta t_{ij}), \quad (5)$$

onde a intensidade do acoplamento entre os neurônios pré (*j*) e pós-sináptico (*i*) é acrescida do produto de uma taxa de modificação sináptica *δ = 0,001 e W* (apresentado em (3)).

### 4.1 Sincronização

A história que acompanha o fenômeno de sincronização de sistemas data de meados de 1665, quando Christian Huygens inventou o relógio de pêndulo[14] e observou o que dois relógios dispostos na mesma superfície de contato apresentavam um movimento sincronizado (Figura 6) [40]. Foi então que realizou testes para verificar se a sincronia persistiria. Perturbou a oscilação dos pêndulos e verificou que após algum tempo, a sincronia voltava a acontecer. Após C. Huygens, diversos pesquisadores investigaram a sincronização em vários sistemas, desde as células cardíacas até o piscar de comunidades de vaga-lumes [41-44].

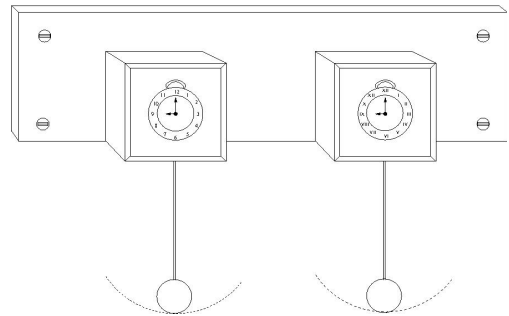

**Figura 6:** Relógios pendulares dispostos na mesma superfície de apoio. São osciladores, onde uma massa é acoplada a um pivô que permite sua movimentação livre.

Em sistemas representados por redes neuronais, é comum observar a sincronização em termos dos picos de disparos ou dos trens de disparos [do inglês *burst*] neuronais. A condição de sincronia entre os neurônios é observada quando os neurônios começam e terminam os trens de disparos simultaneamente [4].

A observação do espaço de parâmetros é auxiliada por uma ferramenta de diagnóstico para a sincronização, o parâmetro de ordem de Kuramoto (*R*). Em 1975, Yoshiki Kuramoto trabalhou com um conjunto de osciladores acoplados todos com a mesma intensidade. Realizando testes com as frequências do sistema, Kuramoto concluiu que os osciladores acoplados ficavam em um círculo unitário no plano complexo [45, 46] e então propôs a equação que corresponde a média da norma do vetor complexo ao longo do tempo

$$R = \langle |\frac{1}{N} \sum_j e^{i\varphi_j}| \rangle, \quad (6)$$

onde $\varphi_j$ é a fase do sítio $j$ representada por

$$\varphi_j = 2\pi \frac{t - t_j(m)}{t_j(m+1) - t_j(m)}, t_j(m) \leq t < t_j(m+1), \quad (7)$$

sendo $t_j(m)$ e $\varphi_j$ o tempo de disparos e fase do neurônio $j$.

---
[14] O relógio pendular de Christian Huygens está exposto juntamente com uma cópia do livro "Horologium Oscillatorium sive de motu pendulorum" no museu Boerhaave na Holanda.

O regime de sincronização obedece uma escala numérica onde o valor máximo do parâmetro de ordem de Kuramoto é 1. No caso deste trabalho, à medida que $R$ aproxima-se de 1, permite diagnosticar mais sincronia entre os neurônios, e a sincronização total seria $R=1$.

## 5 Resultados e discussões

Afim de verificarmos o comportamento temporal da intensidade das conexões em uma rede neuronal com STDP, plotamos a Figura 7. Nesta, verificamos a influência da intensidade da perturbação externa na evolução temporal da intensidade de acoplamento médio entre os neurônios da rede. Os valores iniciais de $a_{ij}$, foram distribuídos normalmente com média 0,5 e desvio-padrão 0,02. Estes fornecem a intensidade do acoplamento entre o neurônio pré-sináptico $j$ e o pós-sináptico $i$, que são mantidos no intervalo [0,0;0,5]. Os neurônios são globalmente conectados e sem auto-conexões, ou seja, com $a_{ii} = 0$.

No caso onde consideramos a ausência de perturbação externa ($I_{ext} = 0,0$), os valores para a intensidade do acoplamento médio entre os neurônios convergem para 0,25. Neste caso a STDP induz um acoplamento unidirecional, onde as sinapses que partem de neurônios de maior frequência para os de menor frequência tendem a 0,5, devido a potenciação e no sentido contrário tendem a zero.

Para o caso onde a inserção de perturbação com intensidades moderadas (5 < $I_{ext}$ < 35), o acoplamento médio assume valores maiores (<K> > 0,4) comparados com o caso onde $I_{ext} = 0,0$ (Figura 8). Nesta situação, a perturbação atua de forma construtiva na rede neuronal, alterando os tempos de disparos, principalmente dos neurônios com frequências de disparos menores. Este efeito leva o sistema a uma alternância entre a potenciação e a depressão das conexões entre os neurônios. Como o valor absoluto da potenciação é maior do que a da depressão, tem-se um saldo positivo no acoplamento médio para tempos suficientemente grandes. No entanto, nos casos onde a perturbação externa assume valores altos ($I_{ext} > 40$), o sistema passa a apresentar um regime fracamente acoplado (<K> ~ 0,0), e assim, este passa a atuar de maneira destrutiva na rede.

A sincronização dos neurônios é diretamente influenciada pela perturbação externa. Comprovamos essa afirmação com o auxílio da Figura 8, onde temos a influência da inserção de perturbação externa no acoplamento médio entre neurônios. Verificamos que conforme a intensidade da perturbação externa sofre acréscimos, os neurônios tendem ao desacoplamento.

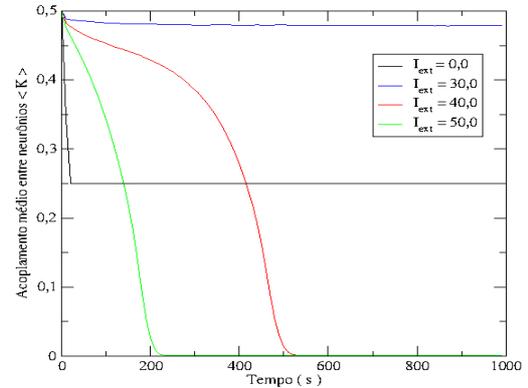

**Figura 7:** Evolução temporal da intensidade média de acoplamento (<K>) para uma rede com 100 neurônios submetidos à diferentes valores da perturbação externa $(\mu A/cm^2)$. A linha em preto representa a ausência de perturbação externa ($I_{ext} = 0,0$), em azul a influência de $I_{ext} = 30,0$; em vermelho $I_{ext} = 40,0$ e em verde $I_{ext} = 50,0$.

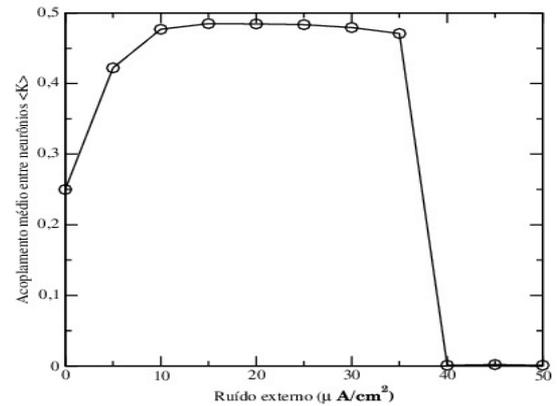

**Figura 8:** Influência da perturbação externa ($I_{ext}$) na intensidade média de acoplamento (<K>) para uma rede com 100 neurônios.

Utilizando o parâmetro de ordem de Kuramoto analisamos o comportamento da sincronização da rede de neurônios acoplados em duas situações: i) sem a inserção do mecanismo da STDP nas equações do modelo e ii) com a inserção da STDP nas equações do modelo. Na Figura 9 observa-se que para o caso sem STDP (linha preta com círculos), conforme aumentamos a intensidade da perturbação externa a sincronização entre os neurônios cai rapidamente. No entanto, na simulação com STDP (linha vermelha com triângulos), para intensidades moderadas de perturbação, o sistema ainda apresenta um alto valor do parâmetro de ordem. Neste caso, a combinação de plasticidade sináptica e perturbação atuam de forma que sistema permanece sincronizado.

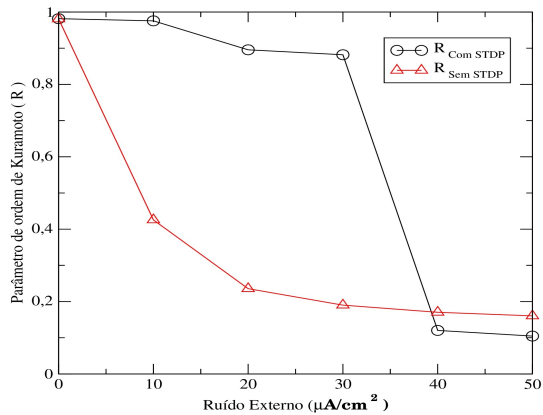

**Figura 9:** Comportamento de sincronia entre neurônios de uma rede com (linha preta com círculos)/sem (linha vermelha com triângulos) plasticidade sináptica (STDP).

## 6 Conclusões

Após estruturar didaticamente a apresentação de um modelo de simulação neuronal, baseado em circuitos elétricos, foi possível investigar a sincronização de disparos em redes neuronais com sinapses modificáveis via STDP mesmo com a presença de perturbação. Observamos que a plasticidade sináptica pode agir como um mecanismo contraposto ao efeito de supressão da sincronização devido à perturbação. Para redes com STDP e intensidades de perturbação moderadas foi observado um acréscimo na intensidade de acoplamento médio entre os neurônios na rede, quando comparados ao caso sem STDP. Neste caso, a combinação de plasticidade sináptica e perturbação atua de forma a reforçar as sinapses pré-existentes. Como consequência a sincronização da rede neuronal torna-se robusta com relação a pertubações externas. No entanto, mesmo com STDP, o sistema evolui para um estado com acoplamento médio fraco, e consequentemente, a um regime não síncrono quando a intensidade da perturbação é suficientemente grande.

## Agradecimentos

## Referências